# Title page

**Title**    Deep Learning-Based Quantitative Assessment of Renal Chronicity Indices in Lupus Nephritis

**Running Title**    DL for Renal CI in LN

**Authors**    Tianqi TU[1,2,3,4#], Hui WANG[11#], Jiangbo PEI[7,8,9#], Xiaojuan YU[1,2,3,4,5], Aidong MEN[9], Suxia WANG[5,6], Qingchao CHEN[7,8*], Ying TAN[1,2,3,4*], Feng YU[2,3,4,10*], Minghui ZHAO[1,2,3,4]

[#]Contributes equally to this work

**Affiliations** 1: Renal Division, Department of Medicine, Peking University First Hospital; Beijing 100034, China

2: Institute of Nephrology, Peking University; Beijing 100034, China

3: Key Laboratory of Renal Disease, Ministry of Health of China; Beijing 100034, China

4: Key Laboratory of CKD Prevention and Treatment, Ministry of Education of China; Beijing 100034, China

5: Renal Pathological Center, Institute of Nephrology, Peking University; Beijing 100034, China

6: Laboratory of Electron Microscopy, Peking University First Hospital; Beijing 100034, China

7: National Institute of Health Data Science, Peking University, Beijing 100191,

China

8: State Key Laboratory of General Artificial Intelligence, Peking University 100191, Beijing, China

9: The School of Artificial Intelligence, Beijing University of Posts and Telecommunications, Beijing 100876, China

10: Department of Nephrology, Peking University International Hospital; Beijing 102206, China

11: Department of Pathology, School of Basic Medical Sciences, Peking University Third Hospital, Peking University Health Science Center, Beijing 100191, China

**Number of text pages**     20

**Number of tables**     2

**Number of figures**     3

**Funding**     National Natural Science Foundation of China under Grant 62201014, the Peking University Medicine Seed Fund for Interdisciplinary Research (BMU2022MX011) and the Fundamental Research Funds for the Central Universities.

**Correspondence:** Qingchao Chen, email: qingchao.chen@pku.edu.cn

Ying Tan, email: tanying@bjmu.edu.cn

Feng Yu, email:  yufengevert1@sina.com

## Abstract

**Background:** Renal chronicity indices (CI) have been identified as strong predictors of long-term outcomes in lupus nephritis (LN) patients. However, assessment by pathologists is hindered by challenges such as substantial time requirements, high interobserver variation, and susceptibility to fatigue. This study aims to develop an effective deep learning (DL) pipeline that automates the assessment of CI and provides valuable prognostic insights from a disease-specific perspective.

**Methods**: We curated a dataset comprising 282 slides obtained from 141 patients across two independent cohorts with a complete 10-years follow-up. Our DL pipeline was developed on 60 slides (22,410 patch images) from 30 patients in the training cohort and evaluated on both an internal testing set (148 slides, 77,605 patch images) and an external testing set (74 slides, 27,522 patch images).

**Results**: The study included two cohorts with slight demographic differences, particularly in age and hemoglobin levels. The DL pipeline showed high segmentation performance across tissue compartments and histopathologic lesions, outperforming state-of-the-art methods. The DL pipeline also demonstrated a strong correlation with pathologists in assessing CI, significantly improving interobserver agreement. Additionally, the DL pipeline enhanced prognostic accuracy, particularly in outcome prediction, when combined with clinical parameters and pathologist-assessed CIs

**Conclusions**: The DL pipeline demonstrated accuracy and efficiency in assessing CI in LN, showing promise in improving interobserver agreement among pathologists. It also exhibited significant value in prognostic analysis and enhancing outcome prediction in LN patients, offering a valuable tool for clinical decision-making.

## Keywords

# Introduction

Lupus nephritis (LN) is a prevalent complication in patients diagnosed with systemic lupus erythematosus (SLE) and exerts a substantial impact on the overall prognosis of the disease.[1,2] Histopathologic assessment of kidney biopsy specimens serves as a pivotal tool in LN, offering significant prognostic and predictive value. The 2018 International Society of Nephrology and Renal Pathology Society (ISN/RPS) classification for LN underscores the inclusion of renal chronicity indices in biopsy reports due to their robust associations with disease progression and ultimate clinical outcomes.[3–5] These indices encompass parameters such as the percentage of sclerotic glomeruli (GS), fibrous crescents (FC), interstitial fibrosis (IF), and atrophic tubules (TA).[3] However, the traditional practice of assessing chronicity indices by pathologists face significant challenges, including substantial time requirements, notable interobserver variability, and susceptibility to fatigue.[6] Therefore, there is a compelling need for innovative techniques to assist pathologists in rapidly and accurately assessing chronicity indices in LN.

The emergence of digitalization in diagnostic pathology has marked a new era of automated biopsy slide analysis. Previous studies have showcased the efficacy of Deep Learning (DL) in augmenting routine histopathologic workflows for pathologists.[7–12] These methods employ deep neural networks to autonomously extract intricate spatial hierarchies of feature representations from whole-slide images (WSIs), utilizing various components such as convolutional layers and transformer structures.[13–19] In the field of renal pathology, prior research has achieved notable success in accurately identifying various histopathologic lesions. This includes distinguishing sclerotic glomeruli and the unified entity of interstitial fibrosis and atrophic tubules (IFTA).[9–12]

However, to the best of our knowledge, no existing research has delved into the specific domain of chronicity indices assessment within LN.

To fill this gap, this study endeavors to develop a robust DL pipeline aimed at automating the assessment of chronicity indices and offering valuable prognostic insights in LN cases. Different from previous methods that employ deep neural networks in an "end-to-end" manner, our DL pipeline is crafted in alignment with the diagnostic reasoning followed by pathologists based on the modified NIH rule.[3] The experiment results indicated that our DL pipeline achieved superior segmentation performance, demonstrated accuracy and efficiency in assessing chronicity indices, and showed promise in improving the interobserver agreement among pathologists. More significantly, in the prognostic experiments, the DL pipeline exhibited significant prognostic value, and demonstrated the potential in enhancing outcome prediction for patients with LN. To the best of our knowledge, this study represents the pioneering and comprehensive exploration of DL approaches for chronicity indices assessment in LN.

## Methods

**A. Patients and Datasets**

This retrospective study involved patients from PKUFH (n=984) and PKU-IH (n=370) between 2010-2013. Figure 1 illustrates the dataset overview. A renal pathologist reviewed cases to confirm that all patients met at least four of the 2019 EULAR/ACR SLE classification criteria at study entry. Tissue sections from both cohorts were prepared with a thickness of 2 μm and stained with methenamine silver and Masson's trichrome (PKUFH: 208 slides; PKU-IH: 74 slides). These slides were digitalized using a Pannoramic Scan whole slide scanner at ×20 magnification (3DHISTECH, Hungary). A data processing procedure is designed to crop WSIs into smaller-sized

(1024×1024 pixels) image patches (appendix p 2) to enable computerized processing, resulting in 105,127 patch images.

30 patients (60 slides, 22,410 patch images) from PKUFH were randomly selected for the training set, while the remaining 74 patients (148 slides) from PKUFH and 37 patients (148 slides) from PKU-IH were designated as internal and external testing sets, respectively. Initial annotation involved six tissue compartments and relevant histopathologic lesions within 22,410 patch images for model training. Annotation was conducted by pathologist WH using the Computer Vision Annotation Tool.[20] Evaluation utilized 7,002 images from the internal testing set and 1,002 images from the external testing set. Chronicity indices were determined independently by three pathologists (WH, YXJ, and WSX) evaluating parameters from whole-slide images. Patient outcomes were assessed based on predefined events: eGFR < 15 ml min$^{-1}$ per 1.73 m$^2$, doubling of serum creatinine, kidney failure with replacement therapy or the last follow-up visit, which took place on December 20, 2023.

**B. Model Development**

The pipeline development comprises three steps: tissue compartment and histopathologic lesion segmentation, diagnostic feature extraction, and chronicity indices calculation (Figure 2). In step 1, three DL models (U-Net and the SAM network[15,19], ResNet[16] and the DeepLabV3$^+$ architecture[18]) were trained to segment six classes: glomerulus, tubule, GS, FC, IF, and TA. Step 2 involves collecting segmentation results and calculating diagnostic features for each patient. Step 3 derives four fully quantitative parameters, leading to the calculation of four semiquantitative chronicity indices sub-scores: GS score, FC score, IF score, and TA score. These sub-scores are then aggregated to obtain the chronicity indices. Further details on model architectures and training methods can be found in the appendix (pp 2–3, 12–13).

**C. Model Validation**

The DL pipeline was tested on the internal testing set (74 patients from PKUFH; 148 slides) and the external testing set (37 patients from PKU-IH; 74 slides). Segmentation performance for glomerulus, tubule, GS, FC, IF, and TA regions was assessed using annotated images from PKUFH (7,002 images) and PKU-IH (1,022 images). The DL pipeline assessed chronicity indices for all patients in internal and external testing sets. Correlation between DL-assessed indices and pathologist-assessed indices was examined. DL-assessed indices were also used to augment interobserver agreement among pathologists. Final chronicity indices were determined through discussions among pathologists and compared with DL-assessed indices.

**D. Statistical Analysis**

We quantitatively evaluated the segmentation performance using Dice's coefficient[21] (referred to as Dice) with 95% Confidence Interval (95% CI). The correlation between pathologists' provided chronicity indices and DL-based chronicity indices was examined by using the Spearman correlation coefficient with p-values ($p \leq 0.05$ was considered statistically significant). The Cohen's kappa coefficient[22] was used to measure the interobserver agreement. For prognostic analysis, Kaplan-Meier method[23] was implemented by using the survminer package in R. CoxPH models[24] were constructed using the survival package in R.

# Results

**A. Demographic and Clinical Characteristics of the Study Cohorts**

Several demographic and clinical features (baseline creatinine, nephrotic-range proteinuria, etc.) were collected at the time of biopsy. Table 1 displays these parameters for all enrolled patients.

Both populations showed similar sex distributions and SLEDAI scores, along with consistent results for most clinical and laboratory assessments. However, differences were noted in age, hemoglobin level, urine protein level, and C3 and C4 levels. Patients from PKUFH were slightly older (mean age: 36.5±11.5 years) compared to those from PKU-IH (mean age: 31.1±11.7 years). PKU-IH patients exhibited lower hemoglobin levels (106.2±22.0 g/L vs. 120.8±14.8 g/L), urine protein levels (4.8±3.8 g/d vs. 3.3±2.9 g/d), C3 levels (0.5±0.2 g/L vs. 0.6±0.2 g/L), and C4 levels (0.11±0.07 g/L vs. 0.13±0.10 g/L). Median follow-up periods were 11.7 years (range: 10.2-12.5 years) for PKUFH and 7.5 years (range: 5.5-11 years) for PKU-IH.

**B. DL–Based Kidney Tissue Segmentation**

Segmentation performance, measured by Dice, is presented in Table 2, with visualizations available in the appendix (pp 15–18). In the internal/external testing set, segmentation performances were as follows: glomerulus (0.938/0.893), tubule (0.924/0.902), GS (0.778/0.734), FC (0.942), IF (0.868/0.813), and TA (0.844/0.802). The averaged Dice across tasks was 0.882/0.837 in the internal/external testing set. Additional segmentation experiments, including ablation experiments and performance with different training data volumes, are detailed in the appendix (pp 9–10, 35–36)

We compared our DL pipeline with state-of-the-art kidney tissue segmentation methods trained in the "end-to-end" manner, including U-Net,[15] DeepLab series,[17,18,25] and Bouteldja's model.[9] Our approach demonstrated superior performance across all segmentation tasks in both internal and external testing sets, as detailed in Table 2. Moreover, we observed that the transformer-based SAM-adapter performed well in segmenting tissue compartments (Glomerulus and Tubule) but exhibited poorer performance in segmenting histopathologic lesions (GS, FC, IF, and TA) within our dataset.

**C. DL-Based Chronicity Indices Assessment**

The DL pipeline demonstrates higher processing speed compared to three pathologists, with potential time savings through improved computer resources (appendix p 19). The Spearman correlation between pathologist(s) and the DL pipeline is shown in Figure 3. The Spearman correlation between pathologists and the DL pipeline in PKUFH was 0.876 ($p<0.0001$), 0.928 ($p<0.0001$), and 0.822 ($p<0.0001$), respectively. In PKU-IH, correlations were 0.862 ($p<0.0001$), 0.909 ($p<0.0001$), and 0.921 ($p<0.0001$), respectively. Detailed correlations for GS/FC/IF/TA scores are available in the appendix (p 20). The DL pipeline exhibits significant correlation with pathologists across these sub-scores, particularly in IF score (internal: 0.950, $p< 0.0001$; external: 0.921, $p< 0.0001$).

The Spearman correlation and the agreement (measured by Cohen's kappa) among pathologists demonstrated improvement with the assistance of the DL pipeline (Figure 3 B1-2, C1-2). Cohen's kappa increased from 0.663 to 0.850 (PKUFH) and from 0.813 to 0.889 (PKU-IH) after using DL. Enhancements in agreement among pathologists for the four sub-scores were notable in both cohorts, except for the TA score in PKU-IH (Cohen's kappa: 0.967 vs 0.967).

## D. Prognostic Experiments

Kaplan-Meier analyses of chronicity indices assessed by the DL pipeline and pathologists on internal and external testing sets are presented in appendix (p 21). We present a more visually accessible result in Figure 3 D-1 and D-2 for easy observation. The DL pipeline provided better FC stratification than pathologists in both cohorts. The utilization of fully quantitative parameters enabled a nuanced assessment of GS/FC/IF/TA, refining the NIH's rule within the range of [0%-50%]. Comparative analyses of scores under these distinct rules indicated the new nuanced scoring rule's statistically significant impact on hazard of the event, particularly in GS/FC/IF scores.

(appendix pp 22–29)

The results of the outcome prediction task are shown in appendix (p 30). AUC improved from using only clinical parameters (PKUFH: 0.733, 95% CI 0.564-0.902; PKU-IH: 0.796, 95% CI 0.649-0.943) to incorporating deep features provided by the DL pipeline (PKUFH: 0.785, 95% CI 0.633-0.937; PKU-IH: 0.874, 95% CI 0.761-0.987). Optimal AUC was achieved with a combination of clinical parameters, pathologists-assessed chronicity indices, and deep features (PKUFH: 0.822, 95% CI 0.773-0.991; PKU-IH: 0.917, 95% CI 0.828-1.000). (appendix p 34)

# Discussion

In this study, we developed a DL pipeline for automated assessment of chronicity indices in LN, showing strong agreement with experienced pathologists. The DL pipeline significantly enhanced interobserver agreement and demonstrated prognostic ability, offering more detailed pathological indicators than conventional scores. Our study represents a pioneering exploration of DL approaches for chronicity indices assessment in LN.

The DL pipeline achieved precise segmentation of kidney tissue compartments and histopathologic lesions (Figure 2), surpassing previous methods by distinguishing IF regions and TA regions.[7–12] It exhibited state-of-the-art performance in both internal and external testing sets, translating into robust alignment with pathologists' assessments. (Table 2). Of significant importance is the inherent stability and resistance to fatigue exhibited by DL methods, and further efficiency could be gained through increased computing resources (appendix p 19). These advances make our pipeline an effective, consistent tool for chronicity indices assessment, which is promising for offering diagnostic suggestions in underserved areas where access to experienced pathologists is frequently limited.[27–30]

The DL pipeline improved agreement among pathologists for chronicity indices, particularly in PKUFH, where significant enhancements were observed in GS, FC, IF, and TA scores. Prognostic analysis revealed the DL-assessed chronicity indices' capability, with FC score offering better stratification than pathologists' assessments.

A significant finding in our method provided fully quantitative proportion parameters, offering flexibility to apply novel scoring rules beyond conventional guidelines (appendix pp 30–33). [31–32] Outcome prediction experiments demonstrated complementarity between pathologists and the DL pipeline, yielding the highest AUC when combined. (appendix p 34).

Our study has several limitations. First, this study is retrospective, and further clinical validation should be conducted prospectively in large patient cohorts. Second, cases with exceptionally high chronicity indices were limited in this study. Further validation using datasets from other institutions may be necessary to substantiate the broader generalizability of our findings. Moreover, challenges emerged during the annotation process, such as the difficulties in delineating the borders between fibrous areas of the renal interstitium and normal tissue. These ambiguities in the annotation introduced noise into the training of DL models.[33] Hence, future efforts on improving the annotation process are required.

In conclusion, the DL pipeline effectively automates chronicity indices assessment, improving interobserver agreement among pathologists. It holds promise for expediting assessment processes, reducing disparities among pathologists, and providing diagnostic suggestions in underserved regions. Additionally, we demonstrated the prognostic value of DL-assessed chronicity indices, highlighting its potential beyond semiquantitative scores. Further validation in large prospective trials is needed to refine findings and assess the clinical utility of the DL pipeline.

# Tables

## Table 1. Demographic and clinical parameters.

| Clinical Evaluation | PKUFH (n=104) | PKU-IH (n=37) | *p*-value |
|---|---|---|---|
| Sex (male/female) | 15/89 | 6/31 | 0.79 |
| Age, yr, mean±SD | 36.5±11.5 | 31.1±11.7 | 0.02 * |
| SLEDAI, mean±SD | 14.6±4.3 | 12.1±3.3 | 0.11 |
| Fever(noninfectious), no.(%) | 35(33.7) | 10(27.0) | 0.54 |
| Malar rash, no.(%) | 25(24.0) | 8(21.6) | 0.83 |
| Photosensitivity, no.(%) | 18(17.3) | 10(27.0) | 0.23 |
| Oral ulcer, no.(%) | 20(19.2) | 6(16.2) | 0.81 |
| Alopecia, no.(%) | 28(26.9) | 9(24.3) | 0.83 |
| Arthralgia, no.(%) | 41(39.4) | 15(40.5) | >0.99 |
| Serositis, no.(%) | 22(21.2) | 8(21.6) | >0.99 |
| Neurological disorder, no.(%) | 9(8.7) | 3(8.1) | >0.99 |
| Nephrotic Syndrome, no.(%) | 71(68.3) | 30(81.1) | 0.41 |
| **Laboratory Assessment** | **PKUFH (n=104)** | **PKU-IH (n=37)** | ***p*-value** |
| Anemia, no.(%) | 72(69.2) | 27(73.0) | 0.83 |
| Acute Renal Failure, no.(%) | 18(17.3) | 4(10.8) | 0.44 |
| Leukocytopenia, no.(%) | 40(38.5) | 13(35.1) | 0.84 |
| Thrombocytopenia, no.(%) | 29(27.9) | 9(24.3) | 0.83 |
| Hematuria, no.(%) | 70(67.3) | 28(75.7) | 0.41 |
| Leukocyturia, no.(%) | 56(53.8） | 22(59.5） | 0.57 |
| Hemoglobin, g/L, mean±SD | 106.2±22.0 | 120.8±14.8 | <0.00 * |
| Urine protein, g/24hr, mean±SD | 4.8±3.8 | 3.3±2.9 | 0.02 * |
| Serum creatinine, μmol/L, median±SD | 124.6±115.4 | 110.8±73.7 | 0.99 |
| C3, g/L, median±SD | 0.5±0.2 | 0.6±0.2 | 0.02 * |
| C4, g/L, median±SD | 0.1±0.1 | 0.1±0.1 | 0.04 * |
| Anti-nuclear antibody, no.(%) | 97(93.3) | 32(86.5) | 0.30 |
| Anti-double-stranded DNA antibody, no.(%) | 77(74.0) | 31(83.8) | 0.27 |

SLEDAI: Systemic Lupus Erythematosus Disease Activity Index scores; * denoted the statistically significance.

**Table 2. Segmentation performance of different deep learning methods on internal (PKUFH) and external (PKU-IH) testing sets.**

| Cohort | Method/Task | Glomerulus Segmentation | Tubule Segmentation | GS Segmentation | FC Segmentation | IF Segmentation | TA Segmentation | Avg. |
|---|---|---|---|---|---|---|---|---|
| PKUFH | U-Net | 0.832 (0.775-0.889) | 0.813 (0.728-0.898) | 0.645 (0.563-0.727) | 0.549 (0.462-0.636) | 0.659 (0.639-0.679) | 0.424 (0.346-0.502) | 0.653 |
| | DeepLab V2 | 0.764 (0.700-0.828) | 0.781 (0.727-0.835) | 0.526 (0.440-0.612) | 0.616 (0.510-0.722) | 0.673 (0.605-0.741) | 0.368 (0.282-0.454) | 0.621 |
| | DeepLab V3 | 0.780 (0.695-0.865) | 0.753 (0.713-0.793) | 0.477 (0.371-0.583) | 0.612 (0.515-0.709) | 0.695 (0.658-0.732) | 0.473 (0.445-0.501) | 0.631 |
| | DeepLab V3+ | 0.786 (0.761-0.811) | 0.791 (0.755-0.827) | 0.549 (0.452-0.646) | 0.595 (0.495-0.695) | 0.730 (0.650-0.810) | 0.509 (0.403-0.615) | 0.660 |
| | Bouteldja's | 0.814 (0.743-0.885) | 0.782 (0.700-0.864) | 0.623 (0.516-0.73) | 0.577 (0.478-0.676) | 0.629 (0.548-0.71) | 0.449 (0.422-0.476) | 0.646 |
| | SAM-adapter | 0.890 (0.791-0.989) | 0.808 (0.746-0.87) | 0.426 (0.322-0.53) | 0.497 (0.416-0.578) | 0.453 (0.426-0.48) | 0.486 (0.433-0.539) | 0.593 |
| | Ours | 0.938 (0.910-0.966) | 0.924 (0.897-0.951) | 0.778 (0.727-0.829) | 0.942 (0.920-0.964) | 0.868 (0.818-0.918) | 0.844 (0.801-0.887) | 0.882 |
| PKU-IH | U-Net | 0.684 (0.648-0.720) | 0.673 (0.632-0.714) | 0.525 (0.498-0.552) | 0.503 (0.466-0.540) | 0.530 (0.431-0.629) | 0.333 (0.303-0.363) | 0.541 |
| | DeepLab V2 | 0.648 (0.548-0.748) | 0.655 (0.570-0.740) | 0.509 (0.469-0.549) | 0.520 (0.473-0.567) | 0.584 (0.528-0.640) | 0.320 (0.232-0.408) | 0.539 |
| | DeepLab V3 | 0.665 (0.620-0.710) | 0.686 (0.599-0.773) | 0.612 (0.522-0.702) | 0.441 (0.380-0.502) | 0.595 (0.547-0.643) | 0.394 (0.314-0.474) | 0.565 |
| | DeepLab V3+ | 0.662 (0.586-0.738) | 0.690 (0.621-0.759) | 0.533 (0.448-0.618) | 0.472 (0.426-0.518) | 0.640 (0.605-0.675) | 0.415 (0.353-0.477) | 0.569 |
| | Bouteldja's | 0.698 (0.610-0.786) | 0.775 (0.609-0.741) | 0.528 (0.458-0.598) | 0.478 (0.409-0.547) | 0.511 (0.437-0.585) | 0.345 (0.329-0.361) | 0.556 |
| | SAM-adapter | 0.826 (0.804-0.848) | 0.891 (0.843-0.939) | 0.417 (0.392-0.442) | 0.497 (0.453-0.541) | 0.525 (0.491-0.559) | 0.470 (0.421-0.519) | 0.604 |
| | Ours | 0.893 (0.873-0.913) | 0.902 (0.880-0.924) | 0.734 (0.604-0.864) | 0.878 (0.789-0.967) | 0.813 (0.728-0.898) | 0.802 (0.746-0.858) | 0.837 |

The performance for different segmentation tasks was reported by Dice (95% CIs). The Avg. represented the averaged Dice over the six segmentation tasks.

# Figures

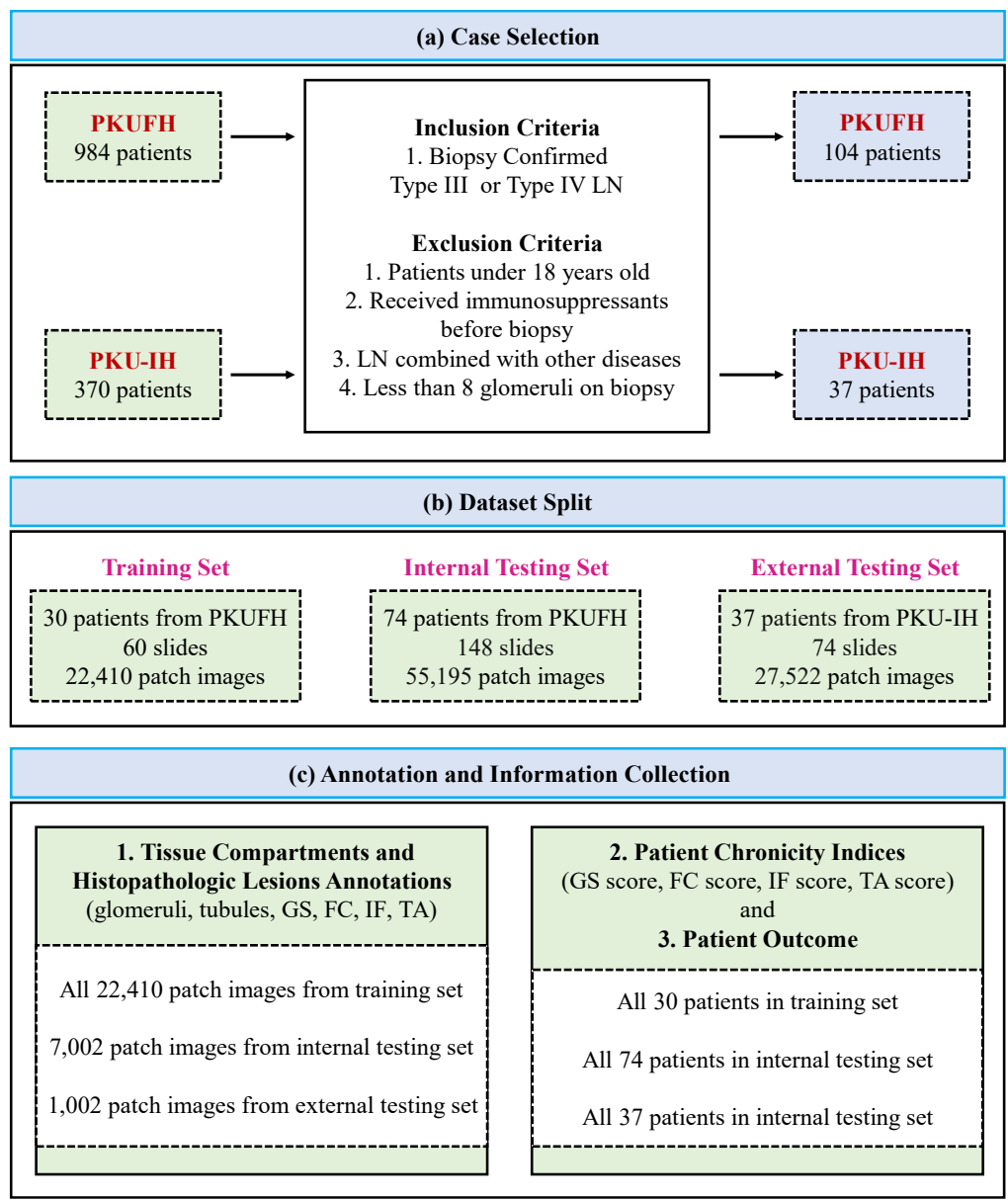

Figure 1. Dataset overview. (a) Data collection process. We curated a dataset consisting of 141 patients with LN across two independent cohorts (PKUFH and PKU-IH) spanning from 2010 to 2013 (b) Dataset split. From the PKUFH cohort, 30 patients (60 slides, comprising 22,410 patch images with size of 1024×1024 pixels) were randomly assigned to the training set. Other 74 patients (148 slides, comprising 55,195 patch images) from PKUFH were as the internal testing set, and 37 patients (74 slides, comprising 27,522 patch images) from the PKU-IH were as the external testing set. (c) Annotation and information collection process. One pathologist manually annotated the cortical regions that related to the chronicity indices assessment (including the glomerulus, tubule, GS, FC, IF, and TA regions), within all images from the training set and randomly selected images from the internal and external testing sets. The chronicity indices of these patients were collected with the help of three expert pathologists. The outcome information for all patients was also collected for prognostic analysis.

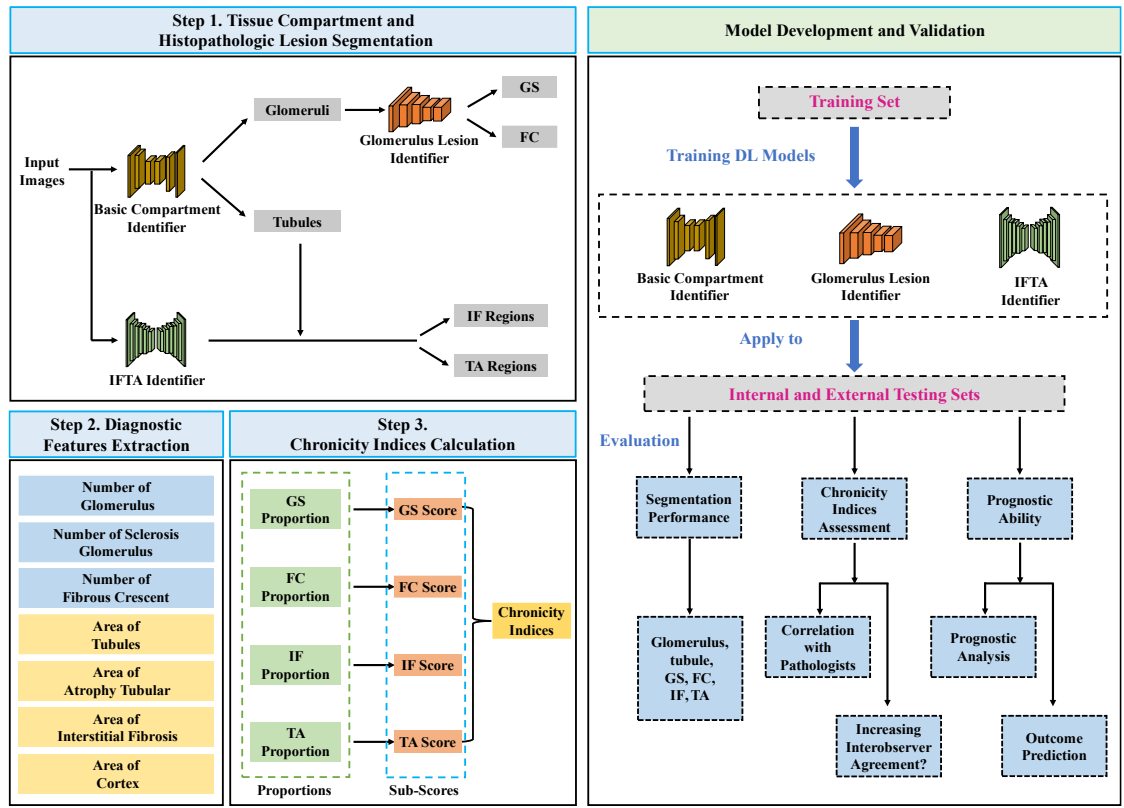

Figure 2. Overall study design: The left illustrates the architecture of the DL pipeline. Step 1 involves integrating three DL models to segment tissue compartments and histopathologic lesions, including glomerulus, tubule, GS, FC, IF, and TA regions. Step 2 extracts diagnostic features for each patient based on segmentation results. Step 3 estimates four proportion parameters from these features, which are then converted into four sub-scores to generate chronicity indices. The right depicts model development and validation. Annotated images from PKUFH (22,410) were used for DL model training with 10-fold cross-validation. The pipeline was applied to a testing set of 74 PKUFH patients and 37 PKU-IH patients. Evaluation focuses on three aspects: segmentation performance, quality of DL-assessed chronicity indices compared with pathologists-assessed indices, and DL pipeline's impact on interobserver agreement among pathologists. Prognostic ability is investigated using prognostic analysis and outcome prediction.

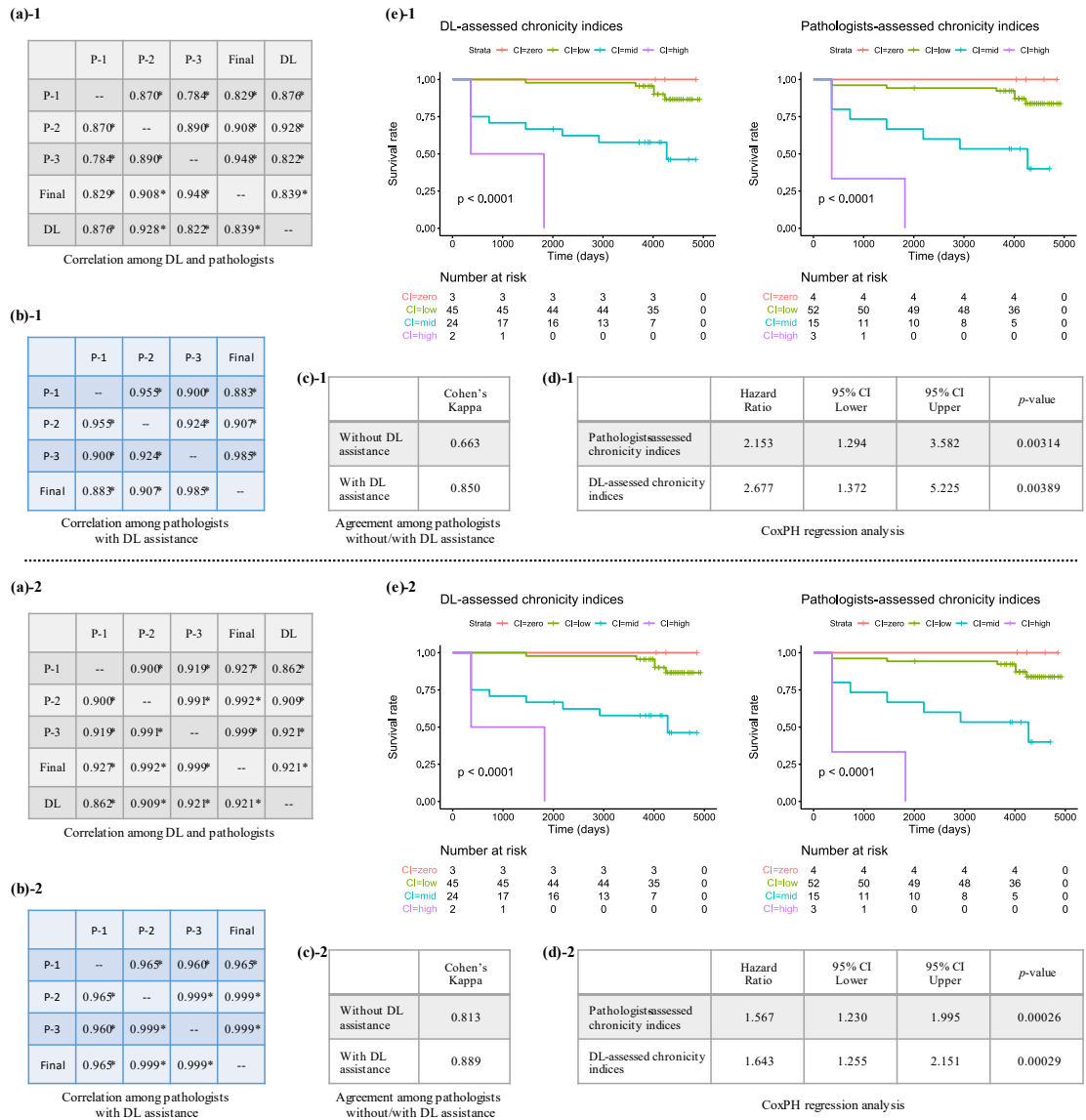

Figure 3. Evaluation in the internal testing set PKUFH (a), and the external testing set PKU-IH (b): (a-1) Spearman correlation between pathologists and the DL pipeline in chronicity indices assessment in PKUFH. P-1, P-2, and P-3 denoted the chronicity indices assessed by pathologist YXJ, WH and WSX. "Final" represented the final chronicity indices determined by the three pathologists. (*: $p<0.0001$). (b-1) Spearman correlation among pathologists with DL pipeline assistance in PKUFH. (c-1) Comparison of agreement (Cohen's kappa) among pathologists with/without DL pipeline assistance in PKUFH. (d-1) Multivariate Cox proportional hazards regression analysis for pathologists-assessed and DL-assessed chronicity indices in PKUFH. (e-1) Kaplan-Meier estimates of survival rate by pathologists-assessed and DL-assessed chronicity indices in PKUFH. (a-2) Spearman correlation between pathologists and the DL pipeline in chronicity indices assessment in PKUIH. (b-2) Spearman correlation among pathologists with the assistance of the DL pipeline in PKUIH. (c-2) Comparison of agreement (Cohen's kappa) among pathologists with/without DL pipeline assistance in PKUIH. (d-2) Multivariate Cox proportional hazards regression analysis for pathologists-assessed and DL-assessed chronicity indices in PKUIH. (e-2) Kaplan-Meier estimates of survival rate by pathologists-assessed and DL-assessed chronicity indices in PKUIH. Chronicity indices were roughly stratified into four categories for visibility. Complete chronicity indices Kaplan-Meier curve is available in the appendix (p 21).

# Acknowledgements

This work is supported in part by the National Natural Science Foundation of China under Grant 62201014, the Peking University Medicine Seed Fund for Interdisciplinary Research (BMU2022MX011) and the Fundamental Research Funds for the Central Universities. The authors thank all the patients enrolled in the study.

# Author contributions

TT, JP, HW, QC contributed to the study design. TT and HW performed the data analysis and were responsible for data collection. JP, HW and TT contributed to the data interpretation and writing of the manuscript. QC, FY, AM, TY and MZ were the senior supervisors of the project. HW, XY and SW curated pathological examinations and assessed chronicity indices. HW and TT had access to and verified the data supporting this study. All authors reviewed and approved the submitted manuscript.

# Conflict of Interest

All authors declare no competing interests.

# Data Availability Statement

All data are available through all corresponding authors upon reasonable request. Codes are available at https://github.com/SPIresearch/DLfochronicity indices.

# Ethics Approval and Consent to Participate

The Institutional Review Board of Peking University First Hospital approved the study (2023[1280]). Informed consent was obtained from all subjects. All experiments were performed in accordance with the Declaration of Helsinki.